\documentclass[runningheads]{llncs}
\usepackage[T1]{fontenc}
\usepackage{graphicx} 
\usepackage{url}
\usepackage{booktabs}
\begin{document}

\title{Towards LLM Accelerated Rapid Reviews for Software Tool Discovery - Case for Log Anomaly Detection}
\titlerunning{Towards LLM Accelerated Rapid Reviews}
\author{Jesse Nyyssölä\inst{1}\orcidID{0009-0006-7276-5696} \and
Hamza Bin Mazhar\inst{1}\orcidID{0009-0009-6352-9810} \and
Alexander Bakhtin\inst{2}\orcidID{0000-0003-3513-7253}
\and
Matteo Esposito\inst{2}\orcidID{0000-0002-8451-3668}\and
Nana Reinikainen\inst{1}\orcidID{0009-0002-3077-3939}\and
Yuqing Wang\inst{1}\orcidID{0000-0003-0175-005X}
\and
Ying Song\inst{1}\orcidID{0000-0001-9791-1879}
\and
Davide Taibi\inst{3,2}\orcidID{0000-0002-3210-3990}
\and
Mika Mäntylä\inst{1}\orcidID{0000-0002-2841-5879}
}

\authorrunning{J. Nyyssölä et al.}
\institute{
University of Helsinki, Fabianinkatu 33, Helsinki, Finland 
\email{\{jesse.nyyssola,hamza.mazhar,nana.reinikainen,yuqing.wang,\\ying.song,mika.mantyla\}@helsinki.fi}
\and
University of Oulu, Pentti Kaiteran katu 1, Oulu, Finland
\email{\{alexander.bakhtin,matteo.esposito\}@oulu.fi}
\and
University of Southern Denmark, Spinderigade 24, Vejle, Denmark
\email{taibi@imada.sdu.dk}
}

\maketitle              

\begin{abstract}
In software engineering research, the primary outcome is frequently a tool. However, for practitioners and academics alike, it is hard to tell which tools are maintained and do they work out of the box. In this paper, we propose a pipeline to identify relevant studies with LLM screening, extract the tools presented in them, and run them with LLM-based coding agent. To evaluate the feasibility of our approach we focus on software log anomaly detection tools. We begin the study by designing a broad search string that yields 3233 hits from Scopus. We request two LLMs to provide an inclusion probability for each title-abstract pair according to the inclusion and exclusion criteria. From the 3233 exported abstracts, this screening reduced the number of included papers to 569, out of which we could download 470. These papers included 206 unique links and after manual evaluation we determined 83 to be tools. Finally, we ran the LLM-based coding agent on these 83 links, and got 24 successfully running tools. As replicating our approach would require roughly only 4 hours of human effort, of which 3 hours were manual PDF downloading, and 12 hours of LLM running time, this demonstrates promising efficiency when utilizing LLMs in rapid reviews. Because practitioner-built tools often lack academic papers, in the future we aim to expand our analysis to tool-hosting platforms such as GitHub and PyPI. In the future, we plan to formalize our workflow as LLM Agent Skills to make our approach easier to adopt.


\end{abstract}

\section{Introduction}
In software engineering research, the primary outcome is frequently a tool \cite{impactofse}. However, utilizing these tools requires  practitioners and academics to search, locate and execute the tools, which is a challenging task given the vast volume of academic literature. This paper introduces an LLM accelerated rapid review methodology to systematically identify and retrieve functional tools from academic publications. 

To understand the feasibility of such a methodology, we test our approach on the area of software log anomaly detection. To be crystal clear, we could have selected any software engineering area where tools are prominent. We chose software log anomaly detection because we possess strong expertise in this topic, while also seeking to ensure that our view of the tools in the area is comprehensive. On a more general level, we further motivate our topic selection as follows.
Software logs contain information about the execution of the system which is used for monitoring, debugging, and maintenance. However, logs can contain millions of lines that has motivated hundreds of academic studies and tool proposals.  
Yet, practitioners have emphasized their preference for tools that can be installed in an hour, are easy to use, and can detect anomalies in five seconds \cite{practitionersexp}. Finding such tools manually can be difficult, which calls for an approach that assesses the tools in a timely manner.

Rapid reviews in software engineering have been endorsed for narrow scoped problem and they are ideal for technology transfer \cite{rrinse}. It has been shown that practitioners are able to apply the outcomes of such reviews \cite{rrinpractice}. Thus, using rapid reviews to identify tools for log-analysis seems like a perfect fit.

Recently, the use of LLMs has been investigated in systematic reviews in software engineering ~\cite{huotalaPromiseChallengesUsing2024a,felizardoChatGPTApplicationSystematic2024,sesreval,petersen2025road,thode2025exploring} 
and in other fields
~\cite{lieberum2025large}.
The consensus is that they make many correct decisions, for example in title-abstract screening \cite{sesreval,lieberum2025large}, although reliability issues remain and their performance has been found similar to that of master's students as early as 2024 \cite{huotalaPromiseChallengesUsing2024a}. Yet, the time reduction provided by LLM-based automated paper screening makes it a promising technique for accelerating rapid reviews.

Currently, it is not only humans who may use software engineering tools but also agentic AI. This further motivates our focus on tool discovery.
For example, Esposito et al. \cite{matteomapek} propose a MAPE-K and agentic AI framework that relies heavily on the continuous ingestion and evaluation of execution logs, making robust, highly scalable tooling a foundational requirement for such agentic architectures. Thus, to realize this vision, we conduct this rapid review of log anomaly detection tools to address the data processing challenges required for successfully automating microservice anomaly detection.

Based on our research objectives, we formulate the following research questions:
\begin{itemize}
   \item RQ1: To what extent can LLMs be used in the rapid review process for software tool discovery?
    \item RQ2: How many tools can we identify and successfully run with our process?
\end{itemize}
The first research question focuses on the process that we are trying to refine, that is, LLM accelerated rapid reviews. By addressing possible strengths or challenges, we provide a substantial academic contribution to future research using similar approaches. The second research question concerns more the output of the pipeline. The set of tools identified by the study is a contribution on its own.

To disambiguate, when we use the term "tool" we refer to software repositories that implement a solution for software log anomaly detection, including replication packages of academic studies. We intentionally do not differentiate between named tools and unnamed replication packages, as the latter may contain functionally equivalent approaches that simply lack a name. In a similar way, when we refer to a repository as successfully running our focus is that it can be installed and executed to produce an output, rather than on its functional correctness or practitioner-readiness (e.g., documentation, containerization).

\section{Methodology}

\begin{figure}[h]
    \centering
    \includegraphics[width=0.6\linewidth]{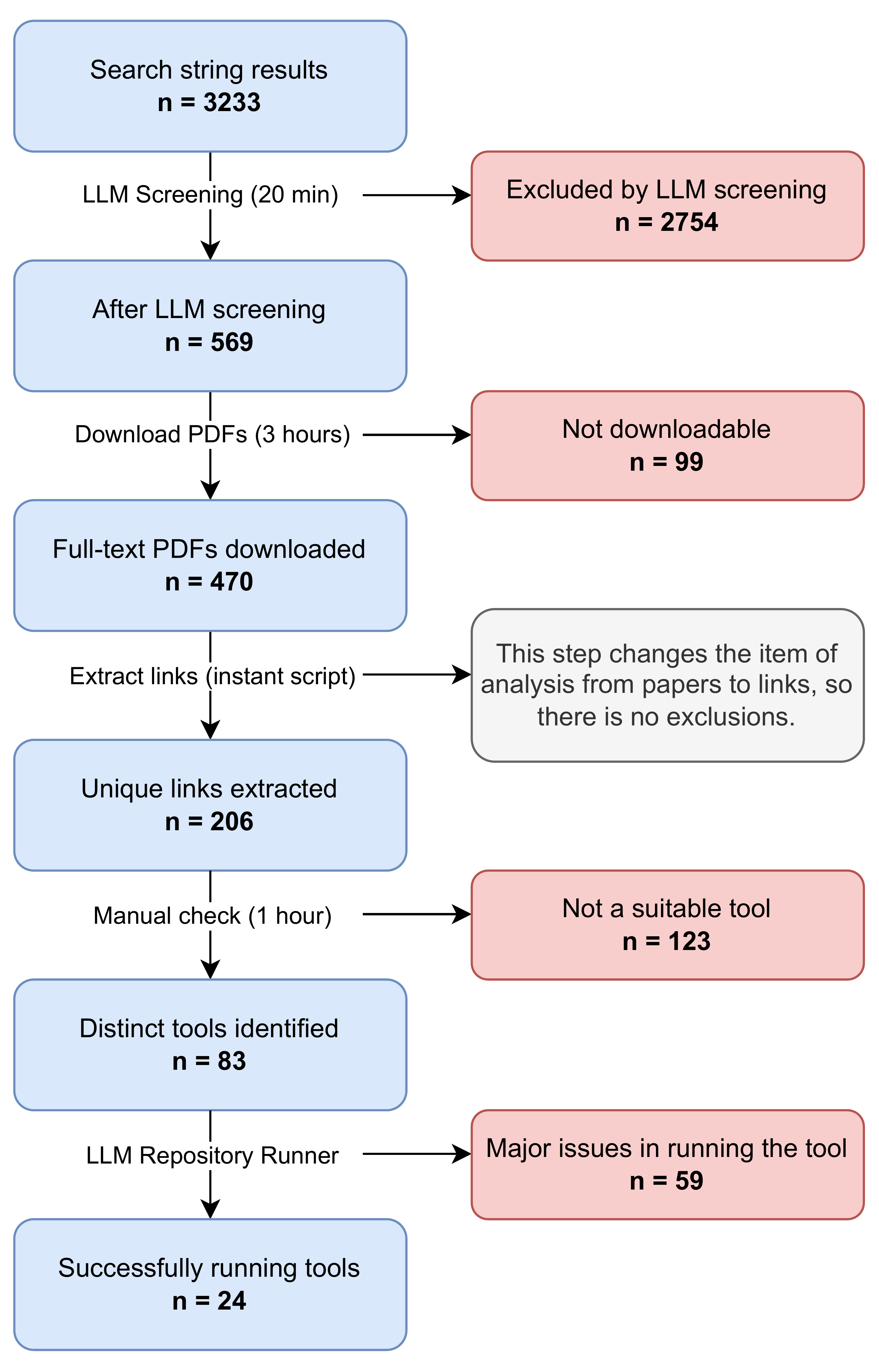}
    \caption{PRISMA-style flow diagram}
    \label{fig:prisma}
\end{figure}

To conduct the study, we designed a pipeline from creating the search string all the way to running the tools with an LLM. This section outlines the methodology and reasoning for our choices. To get an overview of our approach including the results found see Figure \ref{fig:prisma} and for the LLM selection in the different steps see Table \ref{table:llms}. Further details are available in the replication package \cite{rapidreview_replication}.

\begin{table}[h!]
\centering
\caption{LLM Pipeline Configuration}
\begin{tabular}{p{0.16\linewidth} p{0.30\linewidth} p{0.5\linewidth}}
\toprule
\textbf{Stage} & \textbf{Models Used} & \textbf{Rationale} \\
\midrule
\addlinespace[3pt]
LLM Screening & Gemini-2.5-Flash, OpenAI/GPT-4.1-Mini & Past work shows an excellent balance of accuracy, speed, and token cost \cite{sesreval}. \\
\addlinespace[4pt]
LLM Repository Runner & Agent Claude Code (Opus 4.6) & Currently the top-performing coding model for execution tasks \cite{livebench}. \\
\addlinespace[3pt]
\bottomrule
\end{tabular}
\label{table:llms}
\end{table}

\subsection{Search string generation}

In our search, we aim to use a very broad search string that yields a lot of hits in order to ensure we do not miss any studies in key word search phase. Consequently, we formulate the following search string that only focuses on software log anomaly detection.
\begin{quote}
("log anomaly" OR "log-based anomaly" OR "anomaly detection in log*" OR "log analysis") AND ("software" OR "system")
\end{quote}
The search string resulted in 3233 papers as fetched on January 14th 2026 from Scopus, which is one of the largest academic databases containing titles and abstracts from every major publisher like, IEEE, ACM, Springer, Wiley, etc.  

\subsection{LLM abstract screening}
To facilitate abstract screening with LLMs, we use the AISysRev tool \cite{aisysrev}. It is designed to take a Scopus export as a CSV and perform screening with the given inclusion and exclusion criteria. As an output, the tool gives an estimated probability for inclusion as assessed by the chosen model(s). We used two models \textit{gemini-2.5-flash} and \textit{openai/gpt-4.1-mini}. As compact models, they are expected to perform with an excellent balance of accuracy, speed, and token cost. Furthermore, gpt-4.1-mini was the top performing model in large-scale study of title-abstract screening \cite{sesreval}.  

We formulate the following inclusion and exclusion criteria.
\begin{itemize}
    \item IC1 Studies on "Software log analysis"
    \item IC2 Within scope "Anomaly detection on software logs"
    \item EC1 Exclude papers if the target is other than "Anomalies in reliability and testing"
\end{itemize}

The criteria assess the title and the abstract on three key factors: Topic, scope, and target. These criteria only focus on the content of the expected paper rather than the metadata that is handled by Scopus export. We did not want to exclude secondary studies or datasets, because there is a possibility that they present updated scripts or dependencies.
Following the initial LLM screening, we manually validate a stratified sample of 386 abstracts, selected equally from both the included and excluded pools. This sample size, determined using Cochran’s formula~\cite{cochran1977sampling}, provides a 95\% confidence level for evaluating the accuracy of the LLM screening. Two authors evaluate each abstract, with a third author resolving any disagreements.

\subsection{Tool link extraction from full papers}
After title-abstract screening we proceeded to downloading the full papers. Downloading was the most labor intensive part of the whole pipeline as scraping the web page for the download link is generally forbidden in the terms of service of the providers. Our initial test with OpenAlex \footnote{https://openalex.org/} and Unpaywall \footnote{https://unpaywall.org/} APIs could only access roughly 5\% of the papers from our sample pool. This forced us to manually download all papers. 

We assumed that any important tools mentioned in the papers would accompany a link to the tool home page or repository. We recognize that this may not always be the case; however, for this emerging work, it is a practical design limitation we imposed.
For extracting the links from the PDFs, we chose not to leverage LLMs as simple pattern matching for links is very efficient and accurate with regular expressions. Specifically, we created regexes to capture links from GitHub, GitLab, Bitbucket, SourceForge, Zenodo, Figshare, PyPI, and HuggingFace.

\subsection{LLM repository runner}
\label{sec:method_runner}
\begin{figure*}[h]
    \centering
    \includegraphics[width=\textwidth]{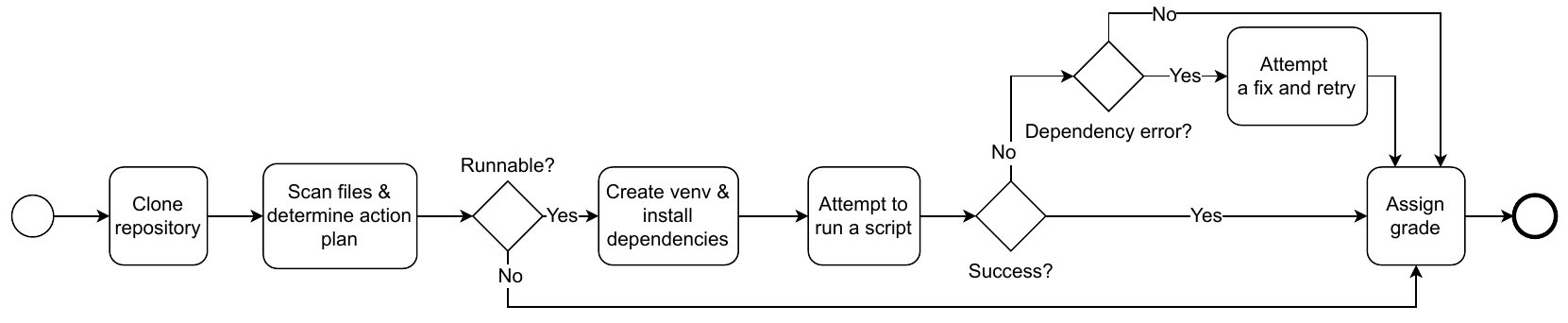}
    \caption{LLM repository runner pipeline}
    \label{fig:llmrunner}
\end{figure*}

Figure \ref{fig:llmrunner} shows the pipeline for the LLM repository runner. The first step of the pipeline is to clone the project, assuming it is a Git link. In case it is other link, such as Zenodo or Huggingface, the LLM agent can inherently handle the necessary downloads. 
In the second step, the agent scans the repository for relevant files for installation such as requirements.txt, pyproject.toml and other setup files. 
In case there are no installation instructions, the agent runs a \emph{grep} command to find import statements. Based on the findings, the agent formulates an action plan that addresses four factors.

\textbf{(1) Is it runnable?} To determine if the tool is runnable, the LLM agent has access to all the files in the repository. We do not define explicit rules for determining if the repository is runnable, it depends on the discretion of the LLM. This filter is intended to address obvious but perhaps unexpected issues, such as no script to run in the first place. In these cases, the repository is simply determined not runnable and the agent ends the process.

\textbf{(2) Data feasibility.} The data on which to run the script can be bundled with the repository, the README might have a link to download it, it could be a common dataset without a link, it can be an unknown dataset without a link, or the data is not mentioned at all. The agent is instructed to handle the three first options, but the last two are left to the discretion of the agent. 

\textbf{(3) Python version.} Python is by far the most popular language used in software log anomaly detection. As such, the prompt focuses on Python related instructions, but at the same time it does not explicitly forbid trying to run non-Python tools.
To determine the correct Python version, the LLM agent utilizes primarily setup or requirement files if they are available. In case the correct Python version is not mentioned, the agent attempts to deduce a suitable version based on the dependencies. 

\textbf{(4) Requirements.} Similar to the Python version, the LLM agent uses available files and information to determine how to install the dependencies. 

Once the action plan is created and the repository determined to be runnable, the pipeline proceeds to creation of a Python virtual environment and installing dependencies. This is done according to the action plan. Once installed, the LLM agent attempts to run an included script. In case it fails due to a dependency issue, the prompt allows a single attempt to fix it as shown in the "Attempt a fix and retry" box of Figure~\ref{fig:llmrunner}. Retrying in this means only attempting to run the script, not the whole pipeline. After that, or due to any other error, the agent is instructed to stop trying to fix the repository.

After trying to run the repository script, the LLM agent is instructed to produce a result file of the outcome. It includes status of the run (success vs. failure), a grading for the installation, and a short assessment. We use the following assessment criteria:
\begin{itemize}
    \item 3 = The tool itself and instructed dependencies worked out of the box, no issues
    \item 2 = Minor issues like dependencies not explicitly given or there was a small issue with them
    \item 1 = Significant issues (e.g., install succeeded but run failed)
    \item 0 = Could not run at all (not runnable, fatal errors, or missing data with no fallback)
\end{itemize}

We utilize \textit{Claude Code agent} with \textit{Opus 4.6} model due to its performance on coding and data analysis related benchmarks \cite{livebench}. It reads a pre-written prompt file that instructs all the steps outlined above. 
To run the setup, we created a dedicated Linux virtual machine with 32 GB of RAM and using 6 cores from Intel Core i7-13700K running on a local computer. 
This setup introduces limitations on what we can successfully run, specifically on GPU operations as it lacks CUDA support. However, we argue that the environment is justified given that it reflects a common use case. Furthermore, there are security considerations when installing unknown repositories and giving the LLM free rein in the system. These factors favor the use of an isolated environment.

\section{Results}

In the results section we outline the key findings of each step in our proposed pipeline. Figure \ref{fig:prisma} shows how the number of included items reduces drastically on each step. Additionally, it contains a rough estimate of how long each step takes to run or perform when assuming that all the data and tools are ready making it essentially reflect the time-cost of re-running our pipeline.

\subsection{LLM title-abstract screening outcomes}
\begin{figure}
    \centering
    \includegraphics[width=0.75\linewidth]{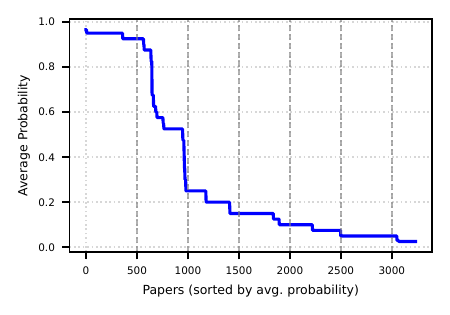}
    \caption{Average inclusion probability of the papers}
    \label{fig:llmscreen}
\end{figure}

Figure \ref{fig:llmscreen} shows the average probability of the two models when the papers are sorted by the average probability. This illustrates well the "elbow" \cite{syakur2018integration} where the probability starts rapidly declining. This is a good indicator of a cut-off threshold for inclusion. We selected the start of the elbow at 0.90 probability as the threshold for inclusion which resulted in 569 included papers out of 3233. 

Human evaluation validates the LLM screening and our "elbow" cutoff point selection. The Cohen's Kappa was 0.839 between the human consensus and the LLM prediction. This signifies almost perfect agreement on Landis \& Koch scale \cite{landiskoch}. When considering the human consensus as the ground-truth, the LLM screening reached precision of 0.933, recall of 0.909 and F1-score of 0.921. 

For our purpose in conducting a rapid review for tool discovery, these are excellent results that clearly meet our expectations as the results are in the top 5\% when compared against SESR-eval results \cite{sesreval}. We think that one reason for such strong results is the data and the criteria, which together form a highly separable set of papers. Many papers are clear includes, even more are clear excludes, and relatively few fall into the ambiguous middle ground. This is evident in Figure~\ref{fig:llmscreen}, where only about 400 papers are in the middle with probabilities between 0.25 and 0.9.

\subsection{Extracted tools from full papers}
From the 569 papers that passed the LLM screening we could download 470. They included 315 links that matched our regexes. There were 206 unique links and 109 duplicates. The most common links in the studies were the repositories for Loghub \cite{zhu2023loghub} (33 hits), Loglizer \cite{he2016experience} (12 hits), and Logdeep \footnote{\url{https://github.com/d0ng1ee/logdeep} The Logdeep repository does not have an associated paper; rather, it is a repository for running some popular approaches.} (10 hits).

We estimated 206 links to be a feasible number for manual assessment to determine if the link contains a suitable tool for our study. What we mean by suitability is that, the link actually leads to a tool for software log anomaly detection. In other words, in this step we exclude all non-relevant tools as well as supplementary links such as datasets without tools. Additionally, we excluded forks. 
The final number of unique links to suitable repositories was 83. 

\subsection{LLM repository runner execution}

Finally, we ran the LLM repository runner as described in Section~\ref{sec:method_runner}. This section starts by introducing performance of the installation process, i.e., how successful it is. In addition, we provide an overview on the execution time distribution across the repositories.

\subsubsection{Install performance}
The number of repositories of each grade is summarized in Table~\ref{tab:grade_distribution}. As shown in the table, the grades 0, 1 and 2 are almost evenly split with most issues caused by dependencies. Unless there are other problems, the extent of the dependency issues appeared to explain the grade to a large degree.

\begin{table}[h]
\centering
\setlength{\tabcolsep}{4pt}
\caption{Repository install grade distribution and common issues per grade}
\label{tab:grade_distribution}
\begin{tabular}{l|c|l}
\hline
\textbf{Grade} & \textbf{\# of repos} & \textbf{Common issues} \\
\hline
Grade 3 & 1 & No issues \\
Grade 2 & 23 & Minor dependency issue \\
Grade 1 & 22 & Missing dependencies, bugs, API incompatibilities \\
Grade 0 & 22 & Repository not available, unresolvable dependencies \\
Timed out & 15 & CPU-bound training, download or processing pipelines \\
\hline
\textbf{Total} & \textbf{83} &  \\
\hline
\end{tabular}
\end{table}

The grade 3 repository was LogLead \cite{mantyla2023loglead}. There were 23 repositories that reached grade 2:  
aecid-alert-aggregation \cite{aminer},
anomaly-detection-log-datasets \cite{criticalreview},
BGL\_AnoDet \cite{nyyssola2024event},
BigLog \cite{biglog},
CAT \cite{catbeyond},
deep-loglizer \cite{deeploglizer},
DeepLog \cite{du2017deeplog},
DistilLog \cite{distillog},
FedLAD \cite{fedlad},
KnowLog \cite{ma2024knowlog},
LL-mod-unsupervised \cite{nyyssola2024speed},
LogDeep,
LogDLR \cite{zhou2025logdlr},
LogEval \cite{logeval},
LoFI \cite{huang2024lofi},
LogGeneratorForAnomalyDetection \cite{loggenerator},
Loglizer \cite{he2016experience},
LogRAG \cite{zhang2024lograg},
Lograph \cite{chu2025lograph},
MoLFI \cite{messaoudi2018molfi},
NeuralLog \cite{le2021neurallog},
OnlineADPipeline \cite{OnlineADPipelineFramework},
and TransSentLog \cite{pham2023transsentlog}.
Note that while the link extraction does not discriminate between primary and secondary studies, this list with its citations was manually collected to try to credit the original authors. 

Table~\ref{tab:grade_distribution} shows that 15 repositories exhausted the 20-minutes execution budget. Most timeouts were attributed to CPU-bound training that progressed too slowly to finish within the budget. Additional timeouts occurred due to prolonged dependency resolutions, heavy dataset downloads, or preprocessing pipelines. In few repositories, the agent could not resolve issues related to external datasets.

All repositories in grade 0 did not execute at all. Most of the repositories were not available, they had either been deleted, were private, or were completely empty. Some had hard unresolvable dependencies like requiring CUDA or external API calls with no alternatives. Some had compatibility issues for libraries and Python versions, while a very few had unresolvable dataset availability issues like missing or proprietary datasets, or broken links. In the 83 repositories, there was only one non-Python project. It depended on Java and got grade 0 because the LLM agent could not install Java.

All repositories in grade 1 showed issues that led to partial or failed executions despite heavy agent intervention. Nearly half of the repositories in this grade lacked any dependency specifications, while the other half had outdated or incomplete specifications. A majority of the repositories also contained code bugs, API incompatibilities, or broken imports. In several cases there were hard-coded CUDA calls with no CPU fallbacks. There were also issues with datasets and preprocessing scripts.

All repositories in grade 2 required minor agent intervention to become runnable. A vast majority of repositories in this grade had incomplete or incorrect dependency specifications, with some repositories completely missing the specifications. For the missing dependencies, the agent had to infer these from import statements within the code. Some repositories required external data or downloading pre-trained model weights adding to the setup overhead. A couple of repositories in this grade had hardcoded CUDA device references which had to be changed to CPU to allow execution. Despite these issues, all repositories produced meaningful output after low-effort troubleshooting from the agent. 

LogLead was the only repository that installed and ran out of the box. One reason that explains LogLead's performance is that it has been maintained after publication to support log analysis research and students who use it in a course.

\subsubsection{Execution times}
\begin{figure}
    \centering
    \includegraphics[width=0.7\linewidth]{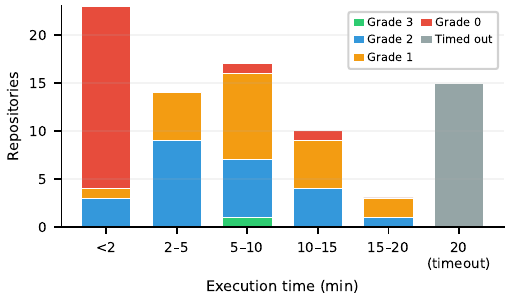}
    \caption{Distribution of the execution times of the repositories}
    \label{fig:execution_times}
\end{figure}

Running the script for all repositories took approximately 12 hours. Figure~\ref{fig:execution_times} shows the distribution of execution times as color-coded by the grade of the repository. As expected, we can observe that the majority grade 0 repositories fail in two minutes. This is due to the LLM agent determining early, as instructed in the prompt, that the repository is unrunnable, which ends the process immediately. If running the repository succeeds in five minutes, most of them have grade 2. Conversely, if it succeeds while taking more than five minutes, the majority switch to grade 1. This is logical, as the LLM agent tries to solve issues within the repository, which both takes time and reduces the grading. 

\section{Discussion}

The LLM worked surprisingly well in various parts of our pipeline. The result of the abstract screening with LLMs was excellent. Trying to get unknown repositories to run by humans is very time-intensive and exhausting, to the point that studies do not do it to this degree. As such, utilizing the LLM can be seen as improving the quality of the review process as a whole. To address \textbf{RQ1}, we argue that the use of LLMs improved both the quality and efficiency of the rapid review process for software tool discovery. For \textbf{RQ2}, we presented a list of 24 repositories that received an installation grade of 2 or 3 by the LLM. Although any qualitative assessment of the tools was outside the scope of this paper, the running tools offer a good starting point for anyone who wants to further assess them or integrate them into agentic frameworks such as MAPE-K \cite{matteomapek}. 

One of the main contributions of the LLM-accelerated rapid review pipeline that we propose is the amount of human effort that is automated. A traditional rapid review of a similar scope would take from days to weeks, but our approach took only around four hours of human effort and 12 hours of running LLMs. Furthermore, majority of the human effort consists of manually downloading PDFs of the papers. While many major publishers provide Text and Data Mining (TDM) API keys for non-commercial research, integrating these into the pipeline would require managing multiple publisher-specific authentication workflows, which would reduce the portability and reproducibility of our approach. Manual downloading is labour-intensive, but it keeps the pipeline self-contained and the results consistent. For scaling this approach to larger reviews, TDM agreements would become justified. Regarding the LLM run time, it is important to note that it took 12 hours to run the 83 repositories sequentially, but outside of rate limits, there is nothing stopping from running several LLM agents in parallel. The 20-minute timeout was chosen both to reflect a realistic practitioner installation scenario \cite{practitionersexp} and to prevent the agent from entering unproductive loops wasting tokens. For further insight for this study, we ran the 15 repositories that timed out again without a time limit and four of them ran well (grade 2), which is a similar ratio as in those that did not time out.

One of our primary metrics for evaluating a tool is whether it executes successfully. This approach may exclude good methods that have poor replication packages. However, given that replicability is a hallmark of science, ensuring the tool runs remains the responsibility of the original authors, not ours. To characterize what predicts successful execution, we collected GitHub metadata for the 49 CPU-compatible repositories and tested associations with grade outcome. Due to the number of repositories, there were no statistically significant results, but some of them were practically meaningful. For example, repositories that included a dedicated \textit{requirements} file ran successfully at a higher rate (61\%) than those without one (38\%). Some factors (README length, number of stars, and number of forks), showed a weak positive trend toward successful execution. Repository age and days since last commit did not predict success.

\section{Threats to Validity}

One notable threat to \textbf{Internal validity} is that it is not always clear whether a failure by the LLM runner is caused by the repository or the agent/prompt. Hence, we risk conflating two different things: How good is the repository and how good is the LLM-agent. Another threat is that we conducted the manual assessment of the links only by a single author which is inconsistent with the paper screening validation, but we reason that it is a much more straightforward process with no controversy. 

The most obvious threat for \textbf{External validity} is that the pipeline was only tested on log anomaly detection tools, and generalization to other SE tool areas is unconfirmed. Another limiting factor is the CPU-only environment where we ran the experiments, and as such we know for certain that the results do not generalize for GPU environments. For better generalizability, using LLM-agents from other providers would also be beneficial. A third threat to external validity, is the potential for evidence loss, i.e., the tools we should have identified. Each instance where a relevant tool is discovered outside our initial search should prompt a critical examination of the reasons, context, and process failures that led to its omission. Another limiting factor is that our pipeline was designed around Python repositories with local dependencies. This does not account for tools which may require Docker containers or cloud infrastructure. However, this is less problematic because research prototypes are often distributed as self contained repositories not utilizing cloud infrastructure, while industrial sources that utilize such infrastructure often do not make their implementation public.

For \textbf{Construct validity}, the main threat is that we assume the tool to be "usable" if the LLM-agent manages to run it. In reality, the usability for humans and LLM-agents differ significantly. For example, for a human adjusting vast configuration files would be very laborious while an LLM could manage it in an instant. Conversely, graphical user interfaces are designed to be intuitive for humans, but an LLM-agent might not be able to run such tools at all. 

There exist threats to \textbf{Conclusion validity} with regards to how the results of this study should be interpreted. They should not be read as saying that only 24 out of 569 log anomaly detection papers provide a good replication package. There are studies in the set that fall outside the scope of tool-oriented research. Additionally, many deep learning based approaches are hardware or time intensive by design. Rather, the 24 tools in 569 papers showcase that in the context of trying to search an easy and fast tool, there are not that many options. However, in this study, we can not guarantee that the running tools produced the correct, expected output. Hence, in terms of output quality of the tools, this study does not intend to place one tool above another.

\section{Future Work}

As the work in this study is highly novel, we will outline future work as an explicit outcome that relates to the limitations of this study which are summarized in Table~\ref{tab:limitations}. The limitations themselves are motivated by the threats to validity presented in the previous section.

\begin{table}[h]
\centering
\caption{Summary of future work by limitation}
\label{tab:limitations}
\small
\begin{tabular}{c @{\hspace{8pt}} p{0.43\textwidth} @{\hspace{8pt}} p{0.43\textwidth}}
\toprule
\textbf{\#} & \textbf{Limitation} & \textbf{Future Work} \\
\midrule
1 &
Tools absent from academic literature (e.g., Angle Grinder) not found.
&
Include additional sources, such as GitHub.
\\[6pt]
2 &
Generalizability: Only evaluated on log anomaly detection tools.
&
Expand to metrics (e.g., CPU, memory) and tracing tools.
\\[6pt]
3 &
Single OS, limited GPU support, and a 20-minute execution timeout.
&
Aim for more robust environments, develop dynamic timeout.
\\[6pt]
4 &
Issue location: Repository or LLM agent?
&
Study LLM runner performance on dataset with groundtruth (e.g., artifact badges).
\\
\bottomrule
\end{tabular}
\end{table}

\textbf{Limitation \#1:} Many significant log analysis tools, such as Angle Grinder\footnote{\url{https://github.com/rcoh/angle-grinder}} and Logfile Navigator\footnote{\url{https://lnav.org}}, are not covered in academic literature and thus fell outside the scope of our rapid review search.
\textbf{Future work:} Extend our approach beyond academic literature to include public hosting platforms (e.g., GitHub) and grey literature. Combining different sources can help produce a more comprehensive picture than running them individually, and the expected overlap between sources serves as a natural cross-validation of the findings.   

\textbf{Limitation \#2:} At this stage we have tested our approach only on tools for software log anomaly detection. Whether our work generalizes to all software tools requires further study.
\textbf{Future work:} We aim to expand the method to other software monitoring tool areas, such as those that analyze metrics (e.g., CPU and memory usage) and those that analyze execution traces from modern microservice systems. Furthermore, we aim to increase the generalizability of the repository runner by formalizing the workflow as Agent Skills. 

\textbf{Limitation \#3:} Limited running environment  with one operating system, limited GPU support, and the 20-minute execution timeout.
\textbf{Future work:} We will explore options to run the repositories in a more robust environment without compromising on information security. Furthermore, the timeout window could be made dynamic if we can identify when the extra time is really necessary (running the tool) as opposed to the LLM agent getting stuck. However, given that practitioners value both ease of use and timely installation \cite{practitionersexp}, the constraints in this study reflect a realistic practitioner scenario, too.

\textbf{Limitation \#4:} While we outlined the several reasons for the LLM runner to fail, it is not always clear if the fundamental issue is in the repository or the agent/prompt.  
\textbf{Future work:} Gather a dataset of studies that have artifact review badges. These can serve as a ground truth so that we can expect the artifact to be runnable. If the LLM still fails, we have a good reason to believe the pipeline is not performing adequately.

\section{Conclusion}
This paper presents an LLM accelerated pipeline for conducting rapid reviews for software tool discovery. We tested our approach on software log anomaly detection tools but we could have chosen any other software engineering area where tools are a typical outcome. We started with a key-word search finding 3233 papers. With the help of LLM abstract screening, regex based link extraction, and finally manual evaluation, we got a list of 83 software tools for log anomaly detection. With the prompt we designed, LLM-based coding agent could eventually run 24 of those with little to no issues.

In the future, we plan to apply our approach to investigate other software monitoring tools. This will provide more evidence of our approach and help make the pipeline more robust. Furthermore, we will extend our approach beyond academic literature for a more comprehensive set of tools. We also plan to formalize our approach as Agent Skills which is an emerging standard for packaging reusable capabilities for LLM agents, as specified in a standard originally developed by Anthropic \cite {AgentSkills2025}.

\begin{credits}
\subsubsection{Data Availability.}
Data is available via our replication package \cite{rapidreview_replication}.

\subsubsection{\ackname}
This work has been funded by the Research Council of Finland (grants n. 359861 and 349488 – MuFAno), and by the Finnish Software Engineering Doctoral Research Network, funded by the Ministry of Education and Culture, Finland.

\subsubsection{\discintname}
The authors have no competing interests to declare that are relevant to the content of this article.
\end{credits}

\bibliographystyle{splncs04}
\bibliography{biblio}

\end{document}